\documentclass[portrait,11pt]{article}
\usepackage[sectionbib,round]{natbib}
\usepackage{times,amsmath,amsfonts}
\usepackage[dvips]{graphicx,graphics}

\usepackage{amsfonts,amsmath,amssymb}
\usepackage{url}
\usepackage{hyperref}
\hypersetup{colorlinks,%
citecolor=blue,%
filecolor=black,%
linkcolor=red,%
urlcolor=blue,%
pdftex}

\newtheorem{theorem}{Theorem}

\newcommand{\bqn}{\begin{eqnarray*}}
\newcommand{\eqn}{\end{eqnarray*}}
\newcommand{\bq}{\begin{eqnarray}}
\newcommand{\eq}{\end{eqnarray}}

\begin{document}
\pagenumbering{gobble}

\title{Functional Data Analysis}
\author{Moo K. Chung \\
University of Wisconsin-Madison\\
\texttt{mkchung@wisc.edu}}
\maketitle

\begin{abstract}
We present a functional data analysis (FDA) framework based on explicit
orthonormal basis expansion for modeling and denoising complex biomedical
signals. Observed functional data are represented as smooth functions in a Hilbert space, and statistical inference is performed directly on their basis
coefficients. This formulation provides a transparent and flexible approach
to smoothing, regularization, and hypothesis testing. Applications to
diffusion tensor imaging tract modeling and EEG denoising demonstrate
the advantages of explicit basis representations for scalable and
interpretable functional modeling.
\end{abstract}

\section{Introduction}

Functional data analysis (FDA) studies data that are 
represented as functions rather than finite-dimensional vectors.
Instead of treating observations as multivariate measurements,
FDA models each sample as a smooth function defined on a continuum
\citep{ramsay.1997,ramsay.2000}. This Hilbert space perspective
provides a principled foundation for smoothing, regression,
principal component analysis, and differential equation modeling. A central idea in FDA is basis expansion: a function $f(t)$ on
$[0,1]$ is represented as a linear combination of known basis
functions, and statistical inference is performed on the coefficient
vectors rather than directly on discretized measurements \citep{chung.2010.SII,wang.2018.annals}.
This separates representation from estimation and enables flexible
regularization and shrinkage.

In this paper, we emphasize explicit orthonormal basis constructions,
in particular the cosine basis obtained from Laplacian eigenfunctions.
Unlike implicit transform-based procedures commonly used in signal
processing, such as the discrete cosine transform, our formulation
is grounded in least squares projection in $\mathcal{L}^2[0,1]$.
This Hilbert space perspective allows direct statistical modeling
at the coefficient level, naturally incorporating shrinkage methods
such as the Wiener filter and stochastic representations through
the Karhunen--Lo\`eve expansion.

The methodology builds on earlier applications of cosine basis
representations to three-dimensional white matter tract modeling
in diffusion tensor imaging \citep{chung.2010.SII} and to EEG
signal denoising \citep{wang.2018.annals}. The resulting framework
integrates projection theory, basis-based least squares estimation,
optimal shrinkage, and stochastic process modeling into a unified
and computationally efficient FDA pipeline. MATLAB codes and sample
data are available at
\url{http://brainimaging.waisman.wisc.edu/~chung/tracts}.

\section{Hilbert space theory in $[0,1]$}
\index{cosine representation}
Consider the $i$-th functional time series data
\bq \zeta_i(t) = \mu_i(t) + \epsilon_i(t) \label{eq:independent},\eq
where $t$ is time variable. We assume the functional data is scaled in such a way that they are defined in $[0,1]$. Formulating the Fourier analysis in a unit interval makes the numerical implementation more convenient.  
$\epsilon_i$ is a zero mean noise at each fixed $t$, i.e., 
$$\mathbb{E} \epsilon_i(t) = 0.$$ 
$\mu_i$ is an unknown smooth functional signal to be estimated. It is reasonable to assume that 
$$\zeta_i, \mu_i \in \mathcal{L}^2[0,1],$$ the space of square integrable functions. Note any function $f \in \mathcal{L}^2[0,1]$ satisfies the condition $$\int_0^1 f^2(t) \; dt < \infty.$$
This condition is needed to guarantee the convergence in the Fourier series expansion. 
Instead of estimating $\mu_i$ in $\mathcal{L}^2[0,1]$, we approximate $\mu_i$ as a linear combination of some known basis functions $\psi_0,$ $\psi_1,$ $\cdots, \psi_k$ by projecting data to the finite subspace spanned by $\psi_0,$ $\psi_1,$ $\cdots, \psi_k$.

Assume $\psi_l$ is orthonormal basis in $[0,1]$ with respect to
the inner product
\bqn \langle f, g \rangle = \int_0^1 f(t) g(t) \; dt.\eqn
Thus, it is required to have
$$\int_0^1 \psi_l (t) \psi_m (t) = \delta_{lm},$$
the Kronecker delta with 1 if $l=m$ and 0 if $l \neq m$. With respect to the inner product, the norm $\| \cdot \|$ is then defined as
$$\| f \| = \langle f, f \rangle^{1/2} = \Big[ \int_0^1 f^2(t) \; dt \Big]^{1/2}.$$
We estimate $\mu_i$ in the  subspace $\mathcal{H}_k$ spanned by up to the $k$ orthonormal basis functions:
$$\mathcal{H}_k =  \Big\{\sum_{l=0}^k c_l \psi_l(t): c_l \in \mathbb{R} \Big\} \subset \mathcal{L}^2[0,1].$$
The least squares estimation (LSE) of $\mu_i$ in $\mathcal{H}_k$ is given by
\bq \widehat {\mu_i } = \arg \min_{f \in \mathcal{H}_k} \big\| f - \zeta_i(t)  \big\|^2. \label{eq:L2min} \eq

\begin{theorem} The minimization of (\ref{eq:L2min}) is given by 
\bq \widehat {\mu_i} = \sum_{l=0}^k \langle \zeta_i, \psi_l \rangle \psi_l,  \label{eq:k-degree}\eq
where the $l$-th degree {\em Fourier coefficient}  $\langle \zeta_i, \psi_l \rangle$ is given by the inner product 
\bq \langle \zeta_i, \psi_l \rangle  =  \int_0^1 \zeta_i(t)\psi_l(t) \; dt. \label{eq:fouriertransform}\eq 
\end{theorem}

{\em Proof.} We need to find function 
$$f(t) =\sum_{l=0}^k c_l \psi_l(t)$$ that is the closest to $\zeta_i$. The distance between $f$ and $\zeta_i$ is given by
$$I(c_0, c_1, \cdots, c_k) = \int_0^1 \Big| \sum_{l=0}^k c_l \psi_l(t) - \zeta_i(t) \Big|^2 \; dt,$$ 
which is a $k+1$ dimensional function in unknown parameter space $(c_0, c_1, \cdots, c_k) \in \mathbb{R}^{k+1}$. Since $I$ is a quadratic function in $(c_0, c_1, \cdots, c_k)$, it has the global minimum at 
$$ \frac{\partial I}{\partial c_0} =  \frac{\partial I}{\partial c_1} = \cdots =  \frac{\partial I}{\partial c_k} =0.$$
The algebraic derivation is straightforward and we have 
$$c_l = \langle \zeta_i, \psi_l \rangle$$
for all $l=0,1, \cdots, k$.   $\square$

The expansion (\ref{eq:k-degree})  is called the {\em Fourier series}. As $k \to \infty$, the expansion converges to $\zeta_i$, i.e.,
$$   \zeta_i (t) = \sum_{l=0}^\infty \langle \zeta_i, \psi_l \rangle \psi_l.$$

\section{Least squares estimation}
\index{parameter estimation}
In practice, functional time series are observed at discrete time points $t_1, t_2, \cdots, t_n$:
\bq \zeta_i(t_j) = \mu_i(t_j) + \epsilon_i(t_j), \quad j=1,\cdots,n \label{eq:discrete}.\eq
The underlying mean functions $\mu_i(t)$ are estimated as 
$$\widehat \mu_i(t) =  \sum_{l=0}^k c_{li} \psi_l(t),$$
where the Fourier coefficients $(c_{0i}, c_{1i}, \cdots, c_{ki})$ for the $i$-th time series is estimated using LSE as follows. At each $t_j$, we have

$$\underbrace{\left( \begin{array}{c} 
\zeta_i(t_1)\\
\zeta_i(t_2)\\
\vdots\\
\zeta_i(t_n) \end{array}\right)}_{Y_i} = \Psi_{n \times (k+1)}  \underbrace{ \left( \begin{array}{c} 
c_{i0}\\
c_{i1}\\
\vdots\\
c_{ik} \end{array}\right)}_{C_i},$$
where the design matrix of basis functions $\Psi$ is given by $$\Psi_{n \times (k+1)} = 
\left( \begin{array}{cccc}
\psi_0(t_1) & \psi_1(t_1)  &\cdots & \psi_k(t_1)\\
\psi_0(t_2) & \psi_1(t_2)  & \cdots & \psi_k(t_2)\\
  \vdots  & \vdots   & \ddots & \vdots \\
\psi_0(t_n) & \psi_1(t_n) &\cdots& \psi_k(t_n)\\
\end{array} \right).$$
The unknown coefficient vector $C_i$ is estimated by multiplying $\Psi ^ \top$ on the both sides:
$$\Psi ^ \top Y_i =  \Psi ^ \top \Psi  C_i.$$
If $n \gg k$, $\Psi ^ \top \Psi$ is full rank and invertible. Thus, we have
\bq C_i = (\Psi ^ \top \Psi)^{-1} \Psi ^ \top Y_i. \label{eq:Ci} \eq
Although we will not show, it is further possible to discretize and reshape the basis $\psi_l$ in such a way that 
$$\Psi^{\top} \Psi = I$$ the identity matrix. The detail is given in \citep{chung.2012.CNA}. This will make the numerical implementation of the Fourier series expansion computationally much more efficient for large $k$. The proposed least squares estimation technique avoids using the often used implicit Fourier transform (FT)  \citep{batchelor.2006,bulow.2004,gu.2004}.

If we have $p$ number of time series, it requires solving equation (\ref{eq:Ci}) $p$ number of times. For really large LSE problems, this is computationally very inefficient. A more efficient way is to estimate the all the coefficients using a single LSE and 
invert $\Psi ^ \top \Psi$ only once by solving

$$Y_{n \times p} = \Psi_{n \times k} C_{k \times p},$$
where
\bqn Y_{n \times p} &=& [Y_1, Y_2, \cdots, Y_p],\\
C_{k \times p} &=& [C_1, C_2, \cdots, C_p] .\eqn
Subsequently, all the coefficients are simultaneously estimated in the least squares fashion as
$$\widehat{C}=(\Psi^{\top} \Psi)^{-1}\Psi^{\top} Y.$$

\section{Wiener filter}
In FDA, observed signals are often contaminated
by observed measurements. After representing the signal in a suitable
orthonormal basis, denoising reduces to estimating the basis
coefficients from noisy observations. A natural criterion for
estimation is the minimization of mean squared error (MSE).
The Wiener filter \citep{wiener.1949} provides the optimal linear estimator under this criterion by shrinking each coefficient according to its
signal-to-noise ratio.

Suppose the observed signal satisfies
\[
y(t) = f(t) + \varepsilon(t),
\]
where $\varepsilon(t)$ is additive noise. In the functional data
analysis framework, we represent the signal using an orthonormal basis
$\{\psi_k(t)\}$ on $[0,1]$:
\[
f(t) = \sum_{k=0}^{\infty} \beta_k \psi_k(t).
\]
Expanding all components in the same basis gives
\[
y(t) = \sum_{k=0}^{\infty} c_k \psi_k(t),
\qquad
\varepsilon(t) = \sum_{k=0}^{\infty} \eta_k \psi_k(t).
\]
Matching coefficients at each frequency yields
\[
c_k = \beta_k + \eta_k.
\]
Assume further
\[
E[\beta_k]=0, \qquad E[\eta_k]=0,
\]
\[
\mathrm{Var}(\beta_k)=\sigma_f^2(k),
\qquad
\mathrm{Var}(\eta_k)=\sigma_\varepsilon^2(k),
\]
and that $\beta_k$ and $\eta_k$ are uncorrelated.

We estimate the signal $f$ by estimating its coefficients $\beta_k$
through linear shrinkage of the observed coefficients:
\[
\widehat{\beta}_k = a_k c_k,
\]
where the shrinkage factor $a_k$ is chosen to minimize the
mean squared error. The mean squared error at frequency $k$ is
\[
R(a_k)
=
\mathbb{E}\big[(\beta_k - a_k c_k)^2\big].
\]
Substituting $c_k=\beta_k+\eta_k$ yields
\[
R(a_k)
=
\mathbb{E}\big[((1-a_k)\beta_k - a_k\eta_k)^2\big].
\]
Because signal and noise are uncorrelated,
\[
R(a_k)
=
(1-a_k)^2\sigma_f^2(k)
+
a_k^2\sigma_\varepsilon^2(k).
\]

Minimizing $R(a_k)$ with respect to $a_k$ gives
\[
a_k
=
\frac{\sigma_f^2(k)}
{\sigma_f^2(k)+\sigma_\varepsilon^2(k)}.
\]

Therefore, the Wiener estimator is
\[
\widehat{\beta}_k
=
\frac{\sigma_f^2(k)}
{\sigma_f^2(k)+\sigma_\varepsilon^2(k)}\, c_k,
\]
and the reconstructed signal becomes
\[
\widehat{f}(t)
=
\sum_{k=0}^{\infty}
\frac{\sigma_f^2(k)}
{\sigma_f^2(k)+\sigma_\varepsilon^2(k)}
\, c_k \psi_k(t).
\]

Thus, in the chosen orthonormal basis, the Wiener filter performs
coefficient-wise shrinkage, scaling each coefficient by the ratio
of signal variance to total variance. This estimator minimizes the
mean squared error among all linear estimators of the form
$\widehat{\beta}_k = a_k c_k$.

\section{Karhunen--Lo\`eve Expansion}

It is often convenient to adopt a stochastic representation that is
more suitable for statistical inference. Assume that the observational
noise $\varepsilon_i(t)$ in (\ref{eq:independent}) is a zero-mean
Gaussian stochastic process on $[0,1]$ with covariance function
\[
C(s,t)
=
\mathrm{Cov}\{\varepsilon_i(s), \varepsilon_i(t)\}.
\]
A stochastic process is simply a collection of random variables indexed
by $t$, and its second-order structure is fully characterized by
$C(s,t)$.

By the Karhunen--Lo\`eve (KL) expansion \citep{adler.1990,dougherty.1999,kwapien.1992,yaglom.1987}, any square-integrable
zero-mean Gaussian process admits the representation
\[
\varepsilon_i(t)
=
\sum_{l=0}^{\infty}
Z_{il}\,\phi_l(t),
\]
where $\{\phi_l\}$ are the orthonormal eigenfunctions of the covariance
operator defined by
\[
\int_0^1 C(s,t)\phi_l(s)\,ds
=
\tau_l^2 \phi_l(t),
\]
and the coefficients satisfy
\[
Z_{il} \sim N(0,\tau_l^2),
\qquad
\mathrm{Cov}(Z_{il},Z_{im}) = 0 \ \text{for} \ l\neq m.
\]
Thus, the KL expansion diagonalizes the covariance operator, and the
random coefficients are uncorrelated and independent under Gaussian
assumptions. In practice, the infinite expansion is truncated to
\[
\varepsilon_i(t)
=
\sum_{l=0}^{K}
Z_{il}\,\phi_l(t) + e_i(t),
\]
where $e_i(t)$ is the truncation error, which becomes negligible when
$K$ is sufficiently large.

If the chosen basis $\{\psi_k\}$ coincides with or approximates the
eigenbasis $\{\phi_l\}$, the stochastic representation becomes
particularly simple and facilitates inference on the coefficient
vectors. The KL expansion therefore provides a probabilistic foundation
for basis representations in functional data analysis.

\section{Cosine basis}

The cosine series representation (CSR) was introduced in \citet{chung.2010.SII}. Unlike the discrete cosine transform used in
signal processing, CSR is formulated through least squares estimation
within a functional data analysis framework.

There are infinitely many possible orthonormal basis in interval $[0,1]$. Here we explain a spectral approach for obtaining orthonormal basis in [0,1]. Consider the space of square integrable functions in $[0,1]$ denoted by $\mathcal{L}^2[0,1]$. Let us solve the eigenfunction
\bq \Delta \psi  + \lambda \psi =0\label{eq:eigen}\eq
in $\mathcal{L}^2[0,1]$ with 1D Laplacian $\Delta = \frac{d^2}{dt^2}$. 
We are basically solving the 2nd order differential equation
\bq \frac{d^2 \psi(t)}{dt^2}  + \lambda \psi(t)  = 0. \label{eq:1Dlaplace}\eq
We can show that eigenfunctions 
$\psi_0,$ $\psi_1,$ $\psi_2,$ $\cdots$ form an orthonormal basis in $\mathcal{L}^2[0,1]$. Note that if $\psi_l$ is an eigenfunction, any multiple of $\psi_l$ is also an eigenfunction. Thus, it is expected the eigenfunctions are properly normalized such that
$$\int_0^1 \psi_l^2 (t) \; dt =1.$$
 The eigenfunctions satisfying (\ref{eq:eigen}) is then given by the usual Fourier sine and cosine basis 
\bq \psi_0(t)=1, \psi_l = \sqrt{2}\sin (l \pi t), \sqrt{2}\cos(l \pi t) \label{eq:sinecosine} \eq
with the corresponding eigenvalues $\lambda_l = l^2 \pi^2$. 
We can check (\ref{eq:sinecosine}) are solutions  by differentiating the eigenfunctions twice.

There are two eigenfunctions corresponding to the same eigenvalue. The multiplicity of eigenfunctions only happen if there is a symmetry in the domain of the Laplace eigenvalue problem (\ref{eq:eigen}). The constant $\sqrt{2}$ is introduced to make the eigenfunctions orthonormal in $[0,1]$.

\begin{figure}
\centering
\includegraphics[width=1\linewidth]{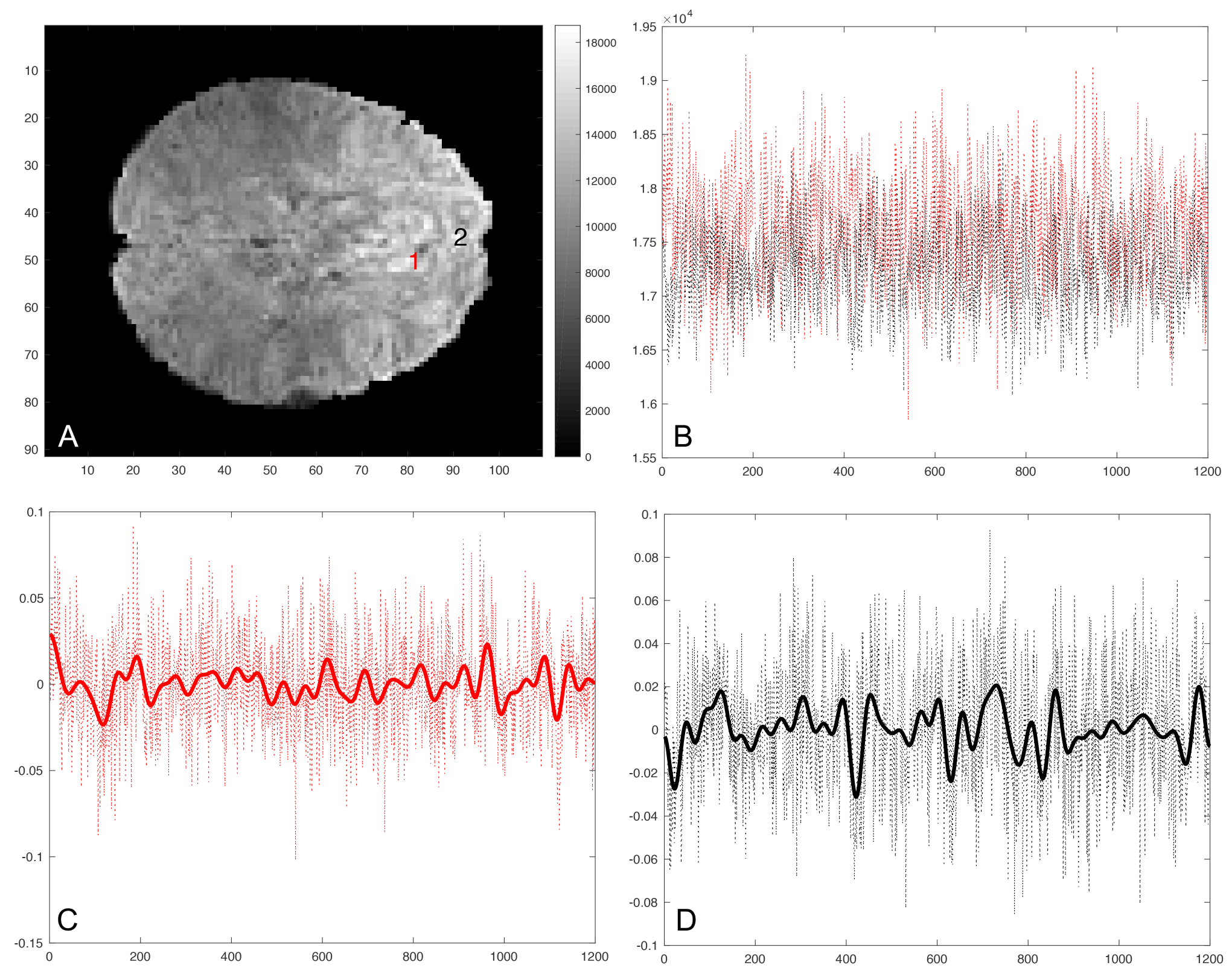} 
\caption{A. fMRI time series at two different voxels at time point $t_1$. B. fMRI time series at voxel 1 (red) and 2 (black) shown for all 1200 time points. C. Normalized and scaled time series at voxel 1 and its cosine series representation with degree $k=59$. D. Normalized and scaled time series at voxel 2 and its cosine series representation with degree $k=59$. Such high-frequency denosising is often necessary for functional signals like fMRI and EEG.}
\label{fig:corr-130114}
\end{figure}

Using both sine and cosine basis is not computationally efficient. Instead of solving (\ref{eq:eigen}) in the domain $[0,1]$, if we add an additional constraint to eigenfunctions, we can get rid of the sine basis. Consider solving the problem in the larger unbounded domain $\mathbb{R}$ with the periodic constraint \bq \psi(t +2) =\psi(t)\label{eq:periodic}.\eq

Suppose $\psi$ is the solution in $[0,1]$. The period 2 constraint forces $\psi$ to be only valid in the intervals $\cdots,$ $[-2, -1],$ $[0,1],$ $[2,3], \cdots$. There are gaps in $\cdots, (-1,0),$ $(1,2),$  $(3,4), \cdots$, where there is no solution. 
We can fill the gap by padding with some arbitrary function. However, if we pad the gaps with any function, it may result in  the Gibbs phenomenon (ringing artifacts)  at the boundary of the intervals $\cdots, 2, 1, 0, 1, 2, \cdots$ \citep{chung.2007.TMI}. To avoid the Gibbs phenomenon, we force the function to be continuous at the boundary  by putting the additional constraint of evenness, i.e., 
\bq \psi(t)=\psi(-t). \label{eq:even}\eq
If $\psi(t)$ is the eigenfunction well defined in  $\cdots,$  $[-2, -1],$ $(-1, 0)$ $[0,1], (1, 2), [2,3], \cdots$ 
we must have   $$\cdots, \psi(t-2)  ,\psi(-t), \psi(t), \psi(-t+2),  \psi(t+2), \cdots$$ 
The only eigenfunctions satisfying  the two constraints (\ref{eq:periodic}) and (\ref{eq:even}) are the cosine basis 
\bqn \psi_0(t)=1, \psi_l (t) = \sqrt{2} \cos (l \pi t)  \label{eq:cosbasis} \eqn
with the corresponding eigenvalues $\lambda_l = l^2 \pi^2$ for integers $l >0$. Then using the cosine basis only, any $f \in \mathcal{L}^2[0,1]$ can be represented as
$$f(t) = \sum_{l=0}^k c_l \psi_l (t) + \epsilon(t),$$
where $c_l$ is the Fourier coefficients and $\epsilon$ is the residual error for using only $k$-th degree expansion. 

If we put the constraint of oddness, i.e., $\psi(t) = - \psi(-t),$
we have sine basis
\bqn \psi_l (t) = \sqrt{2} \sin (l \pi t)  \label{eq:sinebasis}.\eqn

\section{Acknowledgment}
I would like to thank Professor James Ramsay of McGill University for introducing me to this wonderful field of functional data analysis when I was a student there. This study was supported by NIH grants R01 EB022856 and R01 EB028753, and NSF grant MDS-2010778. 
\bibliographystyle{plainnat}

\begin{thebibliography}{14}
\providecommand{\natexlab}[1]{#1}
\providecommand{\url}[1]{\texttt{#1}}
\expandafter\ifx\csname urlstyle\endcsname\relax
  \providecommand{\doi}[1]{doi: #1}\else
  \providecommand{\doi}{doi: \begingroup \urlstyle{rm}\Url}\fi

\bibitem[Adler(1990)]{adler.1990}
R.J. Adler.
\newblock \emph{An Introduction to Continuity, Extrema, and Related Topics for
  General Gaussian Processes}.
\newblock IMS, Hayward, CA, 1990.

\bibitem[Batchelor et~al.(2006)Batchelor, Calamante, Tournier, Atkinson, Hill,
  and Connelly]{batchelor.2006}
P.G. Batchelor, F.~Calamante, J.D. Tournier, D.~Atkinson, D.L. Hill, and
  A.~Connelly.
\newblock Quantification of the shape of fiber tracts.
\newblock \emph{Magnetic Resonance in Medicine}, 55:\penalty0 894--903, 2006.

\bibitem[Bulow(2004)]{bulow.2004}
T.~Bulow.
\newblock {Spherical diffusion for 3D surface smoothing}.
\newblock \emph{IEEE Transactions on Pattern Analysis and Machine
  Intelligence}, 26:\penalty0 1650--1654, 2004.

\bibitem[Chung(2012)]{chung.2012.CNA}
M.K. Chung.
\newblock \emph{Computational Neuroanatomy: The Methods}.
\newblock World Scientific, Singapore, 2012.

\bibitem[Chung et~al.(2007)Chung, Dalton, Shen, Evans, and
  Davidson]{chung.2007.TMI}
M.K. Chung, K.M. Dalton, L.~Shen, A.C. Evans, and R.J. Davidson.
\newblock Weighted {Fourier} representation and its application to quantifying
  the amount of gray matter.
\newblock \emph{IEEE Transactions on Medical Imaging}, 26:\penalty0 566--581,
  2007.

\bibitem[Chung et~al.(2010)Chung, Adluru, Lee, Lazar, Lainhart, and
  Alexander]{chung.2010.SII}
M.K. Chung, N.~Adluru, J.E. Lee, M.~Lazar, J.E. Lainhart, and A.L. Alexander.
\newblock Cosine series representation of 3d curves and its application to
  white matter fiber bundles in diffusion tensor imaging.
\newblock \emph{Statistics and Its Interface}, 3:\penalty0 69--80, 2010.

\bibitem[Dougherty(1999)]{dougherty.1999}
E.R. Dougherty.
\newblock \emph{{Random Processes for Image and Signal Processing}}.
\newblock IEEE Press, 1999.

\bibitem[Gu et~al.(2004)Gu, Wang, Chan, Thompson, and Yau]{gu.2004}
X.~Gu, Y.L. Wang, T.F. Chan, T.M. Thompson, and S.T. Yau.
\newblock Genus zero surface conformal mapping and its application to brain
  surface mapping.
\newblock \emph{IEEE Transactions on Medical Imaging}, 23:\penalty0 1--10,
  2004.

\bibitem[Kwapien and Woyczynski(1992)]{kwapien.1992}
S.~Kwapien and W.A. Woyczynski.
\newblock \emph{{Random Series and Stochastic Integrals: Single and Multiple}}.
\newblock Birkhauser, 1992.

\bibitem[Ramsay(2000)]{ramsay.2000}
J.O Ramsay.
\newblock Differential equation models for satistical functions.
\newblock \emph{The Canadian Journal of Statistics}, 28:\penalty0 225--240,
  2000.

\bibitem[Ramsay and Silverman(1997)]{ramsay.1997}
J.O. Ramsay and B.W. Silverman.
\newblock \emph{Functional Data Analysis}.
\newblock Springer-Verlag, 1997.

\bibitem[Wang et~al.(2018)Wang, Ombao, and Chung]{wang.2018.annals}
Y.~Wang, H.~Ombao, and M.K. Chung.
\newblock Topological data analysis of single-trial electroencephalographic
  signals.
\newblock \emph{Annals of Applied Statistics}, 12:\penalty0 1506--1534, 2018.

\bibitem[Wiener(1949)]{wiener.1949}
N.~Wiener.
\newblock \emph{Extrapolation, interpolation, and smoothing of stationary time
  series: with engineering applications}.
\newblock The MIT press, 1949.

\bibitem[Yaglom(1987)]{yaglom.1987}
A.M. Yaglom.
\newblock \emph{Correlation Theory of Stationary and Related Random Functions
  Vol. I: Basic Results}.
\newblock Springer-Verlag, 1987.

\end{thebibliography}

\end{document}